# Comparative High Pressure Raman Study of Boron Nitride Nanotubes and Hexagonal Boron Nitride

Surajit Saha[1], D. V. S. Muthu[1], D. Golberg[2], C. Tang[2], C. Zhi[2], Y. Bando[2] and A. K. Sood[1*]

[1]*Department of Physics, Indian Institute of Science, Bangalore 560012, India*
[2]*National Institute for Materials Science, Tsukuba, Ibaraki 305-0044, Japan*

**Abstract** High pressure Raman experiments on Boron Nitride multi-walled nanotubes show that the intensity of the vibrational mode at ~ 1367 cm$^{-1}$ vanishes at ~ 12 GPa and it does not recover under decompression. In comparison, the high pressure Raman experiments on hexagonal Boron Nitride show a clear signature of a phase transition from hexagonal to wurtzite at ~ 13 GPa which is reversible on decompression. These results are contrasted with the pressure behavior of carbon nanotubes and graphite.

*Corresponding author: asood@physics.iisc.ernet.in, Fax no. +91-80-23602602

## Introduction

In recent years there have been many experimental and theoretical studies to understand the mechanical properties of carbon single-walled, double-walled and multi-walled nanotubes and their comparisons with each other as well as with the parent compound - graphite. Hexagonal boron nitride (h-BN) is an analogue of semi-metallic graphite with many potential applications like dry lubricants, non-reactive material for nozzles, feed throughs, crucibles and supports in metal, powder metallurgy and glass industries. Very recently, the possibility of preparing boron nitride nanotubes (BNNT) from h-BN sheets, just like preparing carbon nanotubes (CNT) from graphene sheets, has revitalized the interest in BN. The existence of BN nanotubes predicted theoretically [1], was established experimentally by Chopra *et al.* [2] and later the BNNT were synthesized by many groups [3,4]. Due to their novel properties, BNNT are potential materials for applications in various fields. BNNT are far more resistant to oxidation than their counterpart CNT and hence can be used in high temperature applications. BNNT have wide band gap (~5.7 eV), which is very weakly dependent on the chirality and the tube diameter; hence they are promising materials for applications in nano-electronics and optoelectronics. Theoretically it has been predicted [5-7] that BNNT are excellent piezoelectric systems with better response and superior mechanical properties than the usual piezoelectric polymers. On doping BNNT with transition metals, they become half-metallic ferromagnet and hence are potential candidates for spintronics applications, e.g. tunneling magnetoresistance, giant magnetoresistance elements etc. Very recently, large excitonic effects have been observed in BNNT and h-BN [8, 9] which make them promising for applications in optoelectronics as blue light sources and detectors. To understand these potential materials one has to have a complete understanding of the electronic, structural, mechanical and vibrational properties of these materials. Raman spectroscopy under high pressure is one of the established techniques to probe and characterize these properties. In this paper, we report high pressure Raman studies on multi-walled BNNT up to 16 GPa and compare them with the pressure behavior of h-BN crystal carried out upto 21 GPa. We also compare the pressure behavior of BNNT with the known results on carbon nanotubes and graphite.

**Experimental details**

Boron nitride nanotubes were prepared by chemical vapor deposition (CVD) method using carbon free precursors and catalysts similar to Tang *et al.* [10, 11]. The BN nanotubes thus produced were characterized by high-resolution Transmission Electron Microscopy (HRTEM) and Electron Energy Loss Spectroscopy (EELS). The EEL spectrum of the BNNTs indicates that the composition of boron and nitrogen is with a B to N ratio of ~ 1 (±20%). In the BN multi-walled nanotube samples that we have studied, typically there were 5-15 coaxial tubes with an outer diameter ranging from 20-60 nm. The HRTEM images of the BN multi-walled nanotubes are shown in Fig.1. High pressure Raman scattering experiments on BN multi-walled nanotubes were carried out upto 16 GPa and on h-BN upto 21 GPa at room temperature in a gasketed Mao-Bell type diamond anvil cell (DAC) with a mixture of methanol, ethanol and water in the ratio of 16:3:1 as the pressure transmitting medium. The well-known ruby fluorescence technique was used for pressure measurement. The 5145 Å line of an argon ion laser at a power of ~20 mW (at the sample) was used as the excitation source. The scattered photons from the sample were collected in back-scattering geometry and were analyzed using a computer controlled SPEX RAMALOG (double monochromator) spectrometer equipped with a Peltier cooled photomultiplier tube and a photon counting system. The spectral resolution was 5 cm$^{-1}$ and the integration time for each step (step size 0.5 cm$^{-1}$) was 5 s to improve the signal to noise ratio. The Raman frequency and linewidth were obtained by fitting the data to a sum of two Lorentzians – one representing the Raman line from the diamond anvils and the other for the Raman mode of boron nitride.

**Results and Discussions**

Hexagonal BN is a layered structure, like graphite, with B and N atoms arranged in a hexagonal fashion. Two possible ways for packing of BN layers have been suggested, (i) the BN layers are packed just the way graphene sheets are packed in graphite, which has $D_{3h}$ point group symmetry, and (ii) the hexagonal sheets are stacked as alternate layers such that the layer of B atoms comes directly under the layer of N atoms and vice versa, which has $D_{6h}$ point group symmetry. Experiments [12] have shown that the point group symmetry is $D_{6h}$, for which the expected optical modes of vibration at Brillouin zone centre [13] are $A_{2u}$ (IR) + $E_{1u}$ (IR) + $2E_{2g}$ (R) + $2B_{1g}$ ; where IR and R stand for infra-red and Raman active modes. Like in graphite, one of the $E_{2g}$ modes at ~ 52 cm$^{-1}$ corresponds to the shear type rigid layer mode and the other at ~ 1368 cm$^{-1}$ corresponds to the intralayer vibrational mode. For single-walled BNNT, the ab-initio calculations by Wirtz *et al.* [14] predict $3A_{1g}$, $2E_{1g}$ and $3E_{2g}$ Raman active modes. We will discuss the experimental results first on BN multi-walled nanotubes and then h-BN.

**1. BN Multi-walled nanotube**

Raman scattering studies were carried out on BN multi-walled nanotubes under high pressure up to 16 GPa. Raman spectrum of BN multi-walled nanotubes is very similar to that of h-BN. The Raman active $E_{2g}$ tangential mode occurs at ~ 1367cm$^{-1}$ and has a narrow linewidth of ~ 10 cm$^{-1}$. For single-walled BN nanotubes, Wirtz *et al.* [14] have calculated three radial frequency modes below 810 cm$^{-1}$. However, these modes are not expected to be seen in multi-walled nanotubes, as confirmed by our experiments. The variation of the Raman spectra of BN multi-walled nanotubes with different pressures is shown in Fig.2. The intensity of the $E_{2g}$ mode decreases as the frequency increases with increasing pressure and it completely disappears after ~ 12 GPa. The pressure behavior of the Raman frequency of this mode and its linewidth are shown in Fig. 3a and b. The pressure coefficient of the mode frequency is 4.2 cm$^{-1}$/GPa and that of the linewidth is ~ 0.5 cm$^{-1}$/GPa. We increased the pressure upto 16 GPa and then decreased the pressure gradually. Most interestingly, the mode does not recover at all on reducing the pressure to ambient pressure, as seen in the top panel in Fig.2. The fact that the linewidth gets broadened with increasing pressure and the Raman signature disappears, we believe that the BN multi-walled nanotube becomes amorphous after 12 GPa and hence it does not recover on decompression. The pressure effects on carbon multi-walled nanotubes (C-MWNTs) are seen to be reversible upto 20 GPa [15]. X-ray diffraction experiments show that C-MWNTs become irreversibly amorphous after ~ 51 GPa [16]. Therefore, a comparison of BN multi-walled nanotubes with C-MWNTs reveals that BN multi-walled nanotubes are more easily deformable. We will now compare the above results with the pressure behavior of its crystalline counterpart h-BN.

**2. Hexagonal BN**

Very recently, resonance Raman studies on cubic and h-BN crystals have shown that the non-resonant Raman cross-section for hexagonal phase is about 10 times larger than that of cubic phase [17]. Our Raman spectrum of h-BN at ambient pressure shows the high frequency intralayer Raman active $E_{2g}$ mode at ~ 1368 cm$^{-1}$ with a linewidth of ~ 21 cm$^{-1}$, similar to the reported results [17-20]. We could not see the low frequency $E_{2g}$ shear mode at 52 cm$^{-1}$ possibly because of low intensity of the mode and high Rayleigh background. Raman spectra of h-BN at different pressures are shown in Fig.4 (a) for increasing pressure run. The maximum pressure was ~ 21 GPa. A few typical Raman spectra in the decreasing pressure run are shown in Fig.4 (b). We see that in contrast to BN multi-walled nanotubes, the Raman mode is clearly seen to recover on decompression. The mode frequencies are plotted by filled circles in increasing pressure run and by open circles in decreasing pressure run in Fig.5 (a). As the pressure is increased, the $E_{2g}$ mode hardens with a slope of ~ 4.3 cm$^{-1}$/ GPa, similar to the result obtained by Kuzuba *et al.* [18], with a gradual broadening of the linewidth at a rate of ~ 3 cm$^{-1}$/GPa (Fig.5 (b)). The pressure coefficient of the mode frequency is similar to that in BNNT. However, the increment in linewidth with pressure is much higher in h-BN than that in nanotubes. As shown in Fig.5 (a), the Raman mode frequency abruptly drops at ~

13 GPa and again starts increasing with increasing pressure with a slope of ~ 3.8 cm$^{-1}$/GPa. This change in response of the Raman mode with pressure at 13 GPa is a clear signature of a phase transition, which we believe is a transition from hexagonal (h) phase to the metastable wurtzite (w) phase as has been shown by x-ray experiments [21,22] and first principle DFT calculations [23]. We also see a small hysteresis in the phase transition pressure. This phase transition from h-BN to w-BN at ~ 13 GPa is very similar to the reversible phase transition in graphite from hexagonal to wurtzite phase at ~ 15 GPa. At high temperature (~ 2000$^o$C) and high pressure (~8.5 GPa) h-BN transforms to cubic BN [22,24], the second hardest material next to diamond which itself is a transformed form of graphite at high temperature and high pressure. Apart from the high frequency $E_{2g}$ mode, we have also seen two weak modes at ~ 880 cm$^{-1}$ and ~ 1030 cm$^{-1}$ only under high pressure which could be followed up to ~ 8 GPa with pressure gradients of 4.2 cm$^{-1}$/GPa and 3.9 cm$^{-1}$/GPa respectively. Interestingly, these two modes are also reversible under the pressure cycle. Since there is no Raman active mode in this range for h-BN, these two modes may be disorder-induced Raman modes. Exact origin is not known at present.

**Conclusion**

We have performed high pressure Raman scattering experiments on boron nitride multi-walled nanotubes, which show a contrasting pressure behavior compared to its counterpart carbon multi-walled nanotubes. We show that BN multi-walled nanotubes undergo an irreversible phase transition possibly to an amorphous phase at a much lower pressure of ~ 12 GPa, as compared to ~ 51 GPa in carbon multi-walled nanotubes. This may be due to strong ionic nature of the bonding in boron nitride as compared to covalent bonding in carbon. The starting crystalline hexagonal phases in BN as well as in carbon show a phase transition to wurtzite phase at similar pressures - 13 GPa for h-BN and 15 GPa for graphite. Since the bulk modulus of h-BN is 36.7 GPa [25] and that of graphite is 33.8 GPa [26], h-BN is slightly less deformable than graphite. It is interesting to see that in hexagonal phase both the systems have almost similar mechanical property, but once they are rolled up to make nanotubes, the property becomes quite different. Thus our Raman results suggest that high pressure X-ray experiment on BN multi-walled and TEM on pressure cycled BN multi-walled nanotubes should be interesting to carry out to confirm the pressure induced amorphization after ~ 12 GPa.

**Acknowledgement**

AKS thanks the Department of Science and Technology, India, for financial assistance.

**Figure captions**

**Fig. 1**: HRTEM images of BN multi-walled nanotubes. The inset shows a zoomed portion.

**Fig. 2:** Raman spectra of BN multi-walled nanotubes at various pressures. The Raman signature is irreversible after ~ 12 GPa.

**Fig. 3:** Pressure behavior of Raman frequency and linewidth. The solid line is a linear fit to the data.

**Fig. 4:** Raman spectra of h-BN with increasing (a) and decreasing (b) pressures. The Lorentzian background is because of diamond Raman line and the other Lorentzian is the high frequency $E_{2g}$ mode.

**Fig. 5:** Pressure behavior of Raman mode frequency (a) and its linewidth (b). A signature of h-BN to w-BN phase transition is seen at ~ 13GPa in Raman mode frequency (a) with a small hysteresis.

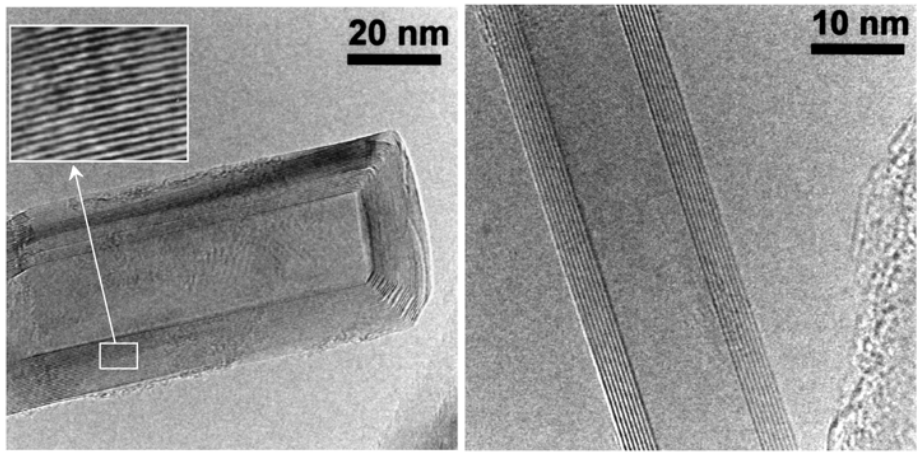

**Figure 1**

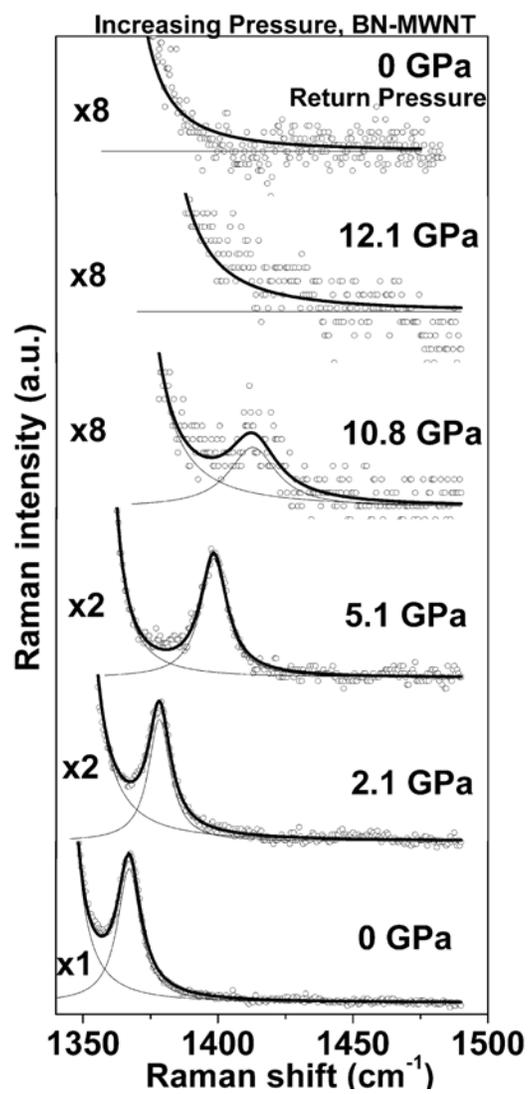

**Figure 2**

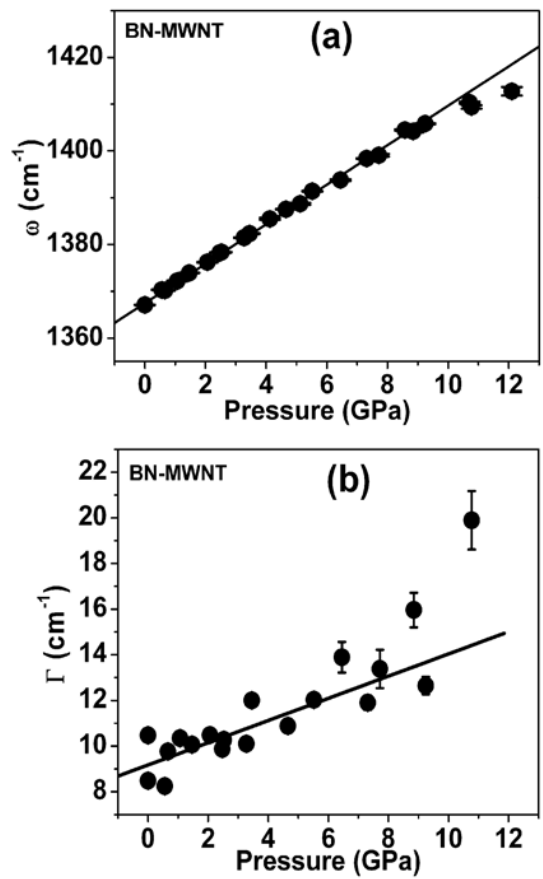

**Figure 3**

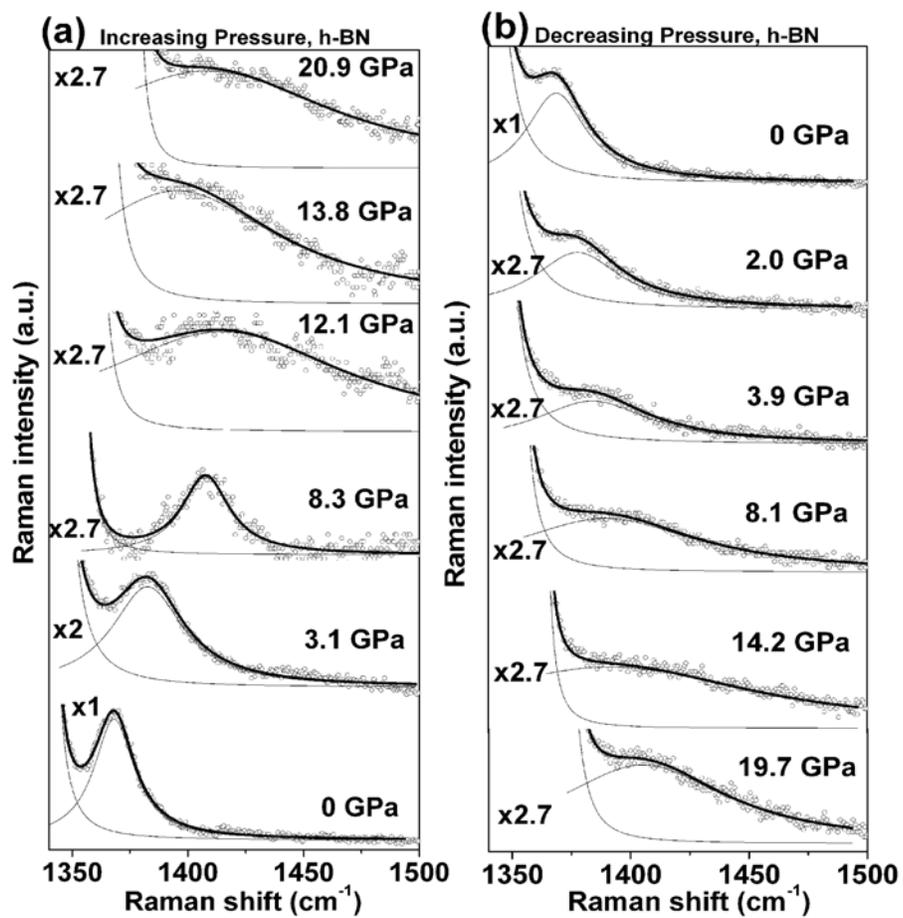

**Figure 4**

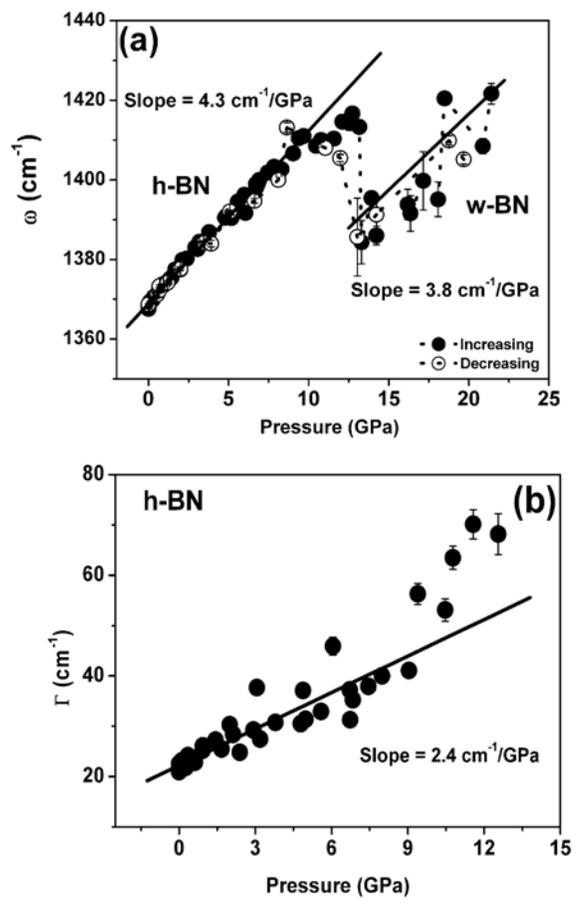

**Figure 5**